\documentclass[aps,prb,twocolumn,showpacs,psfig,superscriptaddress,longbibliography]{revtex4-1}
\usepackage{textcomp}
\usepackage{times}
\usepackage{graphicx}
\usepackage{float}
\usepackage{latexsym,amsmath,amssymb,bm,euscript}
\usepackage{color}
\usepackage{subfigure}
\usepackage{epstopdf}
\usepackage[colorlinks=true,linkcolor=blue,citecolor=blue]{hyperref}
\usepackage{hyperref}
\usepackage{soul}
\usepackage[normalem]{ulem}
\usepackage{mathrsfs}
\usepackage{amsmath}
\usepackage{lettrine}
\usepackage{xspace}
\usepackage{textcomp}
\usepackage{array}
\usepackage{multirow}

\begin{document}

\title{Field-induced antiferromagnetism and Tomonaga-Luttinger liquid behavior in the quasi-one-dimensional Ising-Antiferromagnet SrCo$_2$V$_2$O$_8$}

\author{Yi Cui}
\thanks{These authors contributed equally to this work.}
\affiliation{Department of Physics and Beijing Key Laboratory of
Opto-electronic Functional Materials $\&$ Micro-nano Devices, Renmin
University of China, Beijing, 100872, China}

\author{Y. Fan}
\thanks{These authors contributed equally to this work.}
\affiliation{Department of Physics and Beijing Key Laboratory of
Opto-electronic Functional Materials $\&$ Micro-nano Devices, Renmin
University of China, Beijing, 100872, China}
\affiliation{Institute of Physics, Chinese Academy of Sciences, Beijing, 100190, China}

\author{Z. Hu}
\affiliation{Department of Physics and Beijing Key Laboratory of
Opto-electronic Functional Materials $\&$ Micro-nano Devices, Renmin
University of China, Beijing, 100872, China}

\author{Zhangzhen He}
\affiliation{State Key Laboratory of Structural Chemistry, Fujian Institute
of Research on the Structure of Matter, Chinese Academy of Sciences, Fuzhou,
Fujian 350002, China}

\author{Weiqiang Yu}
\email{wqyu\_phy@ruc.edu.cn}
\affiliation{Department of Physics and Beijing Key Laboratory of
Opto-electronic Functional Materials $\&$ Micro-nano Devices, Renmin
University of China, Beijing, 100872, China}

\author{Rong Yu}
\email{rong.yu@ruc.edu.cn}
\affiliation{Department of Physics and Beijing Key Laboratory of
Opto-electronic Functional Materials $\&$ Micro-nano Devices, Renmin
University of China, Beijing, 100872, China}


\begin{abstract}

We investigate the low-temperature properties of the Ising-like screw chain antiferromagnet SrCo$_2$V$_2$O$_8$
under a longitudinal magnetic field
by susceptibility and $^{51}$V NMR measurements.
The bulk susceptibility $\chi$ shows an onset of
long-range Ising-antiferromagnetic (AFM) order and the suppression of the order by field
with the N\'{e}el temperature dropped from 5.1~K to 2~K when field increases from 0.1~T to 4~T.
The suppression of the AFM order by the field is also observed
by the NMR spectra and the spin-lattice relaxation $1/T_1$.
At fields above 4~T,
$\chi$ shows a low-temperature upturn, which is consistent with the
onset of a transverse antiferromagnetic order
as supported by the quantum Monte Carlo simulations.
A line splitting in the NMR spectra is also observed
at high temperatures.
We show that the line split characterizes the
onset of a short-range transverse antiferromagnetic order
with magnetic moments orientated along the crystalline [110]/[1$\bar1$0]
directions. The $1/T_1$ data at
higher temperature
show a power-law
behavior $1/T_1{\sim}T^{\alpha}$, which is consistent
with the Tomonaga-Luttinger-liquid behavior.
With increasing the field, the power-law exponent $\alpha$
changes from negative to positive,
which clearly shows an
inversion of the Luttinger exponent $\eta$,
where the dominant low-energy spin fluctuations switch from the
longitudinal type to the transverse type at a high field of 7~T.

\end{abstract}

\maketitle

\section{Introduction}

One-dimensional (1D) Ising anisotropic magnets contain strong quantum fluctuations that can be easily
tuned by magnetic field or pressure~\cite{Sachdev,Giamarchi},
and, therefore, offer an ideal platform for exploring rich phases, novel excitations, and quantum phase
transitions. However, quasi-1D Ising magnetic materials
are still rare. Recently,
the spin-$1/2$ Ising antiferromagnets BaCo$_2$V$_2$O$_8$ and SrCo$_2$V$_2$O$_8$ ~\cite{He_PRB_2005,He_PRB_2006} have
attracted a lot of attention because a number of exotic properties, such as
the $E_8$ spinon bound states~\cite{Coldea_Science_2010,Wu_PRL_2014,Zhang_PRB_2020,zou_E8_2020},
string excitations~\cite{Wang_nature_2018,Wang_PRL_2019,Bera_NP_2020}, and novel quantum critical behaviors~\cite{Faure_NP_2018,Cui_PRL_2019}
have been discovered in these two systems.

SrCo$_2$V$_2$O$_8$ crystallizes in the tetragonal $I$4$_1cd$ space group
with screw chains of CoO$_6$ octahedra running along the $c$ axis~\cite{zhangzhenhe_JCG_2006}.
Each unit cell contains four screw chains with antiscrewing direction among the nearest
neighboring chains, which are separated by nonmagnetic Sr$^{2+}$ and V$^{5+}$ ions on the $ab$ plane.
At zero magnetic field, it is
antiferromagnetically ordered at about 5~K with Ising moments pointing along
the $c$ axis. Within each layer of the $ab$ plane, the moments among the diagonal chains are
antiparallel as a consequence of the interchain couplings~\cite{Niesen_PRB_2013,Niesen_PRB_2014,Shen_NJP_2019}.

For an Ising chain under a longitudinal magnetic field, a spin disordered state described by a Tomonaga-Luttinger liquid (TLL),
is induced above a critical field~\cite{Haldane_PRL_1980}.
Meanwhile, the
spin excitation spectrum at the temperature above the ordering temperature
is dominated by a longitudinal mode in
the intermediate magnetic field region above the critical field and further by a transverse mode
at higher fields~\cite{Haldane_PRL_1980,Bogoliubov_NPB_1986}.
In real materials, the interchain couplings are present,
and a longitudinal spin density wave (LSDW) state and a transverse
antiferromagnetic (TAF) state can be stabilized sequentially at finite temperatures with increasing fields.
These phases have been investigated in BaCo$_2$V$_2$O$_8$ and SrCo$_2$V$_2$O$_8$ by elastic neutron scattering
studies where the LSDW phase is shown to exist in a narrow field range from
4~T to ~9 T for BaCo$_2$V$_2$O$_8$~\cite{Canevet_PRB_2013,Grenier_PRB_2015,Faure_PRL_2019} and
from 4~T to 7~T for SrCo$_2$V$_2$O$_8$~\cite{Shen_NJP_2019}, respectively, and the TAF phase
is expected to emerge at higher fields~\cite{Canevet_PRB_2013,Grenier_PRB_2015,Shen_NJP_2019}.

Actually, the high-field phase diagram is still not fully clear for the two compounds. For example,
the string excitations associated with longitudinal spin excitations
have been observed to survive
up to 28.7~T in SrCo$_2$V$_2$O$_8$ by terahertz spectroscopy~\cite{Wang_nature_2018},
which is far beyond the field range where the LSDW phase is stabilized.
Moreover, an NMR study on BaCo$_2$V$_2$O$_8$ reveals a sequence of incommensurate phases with field up to 22.8~T~\cite{Klanjsek_PRB__2015}, whose origin needs to be verified.
Theoretically, BaCo$_2$V$_2$O$_8$ was modeled by a quasi-1D XXZ model~\cite{Okunishi_PRB_2007}.
By using the bosonization technique combined with an interchain mean-field approximation, 
the critical field from N\'{e}el order to LSDW order was calculated to be
consistent with experiments~\cite{Kimura_PRL_2008,Kimura_PRL_2008_2,Canevet_PRB_2013}.
A coexistence of the LSDW and the TAF orders was found
to survive up to 15.1~T, beyond which the TAF order dominates.
But in quantum Monte Carlo calculations, no coexistence of LSDW
and TAF phases was found with Ising-like interchain couplings ~\cite{Fan_PRR_2020}.
To clarify the evolution of phases and corresponding dominant spin excitations with field,
further experimental and theoretical studies are
highly demanded.

In this paper, we report our magnetic susceptibility and $^{51}$V NMR measurements
on SrCo$_2$V$_2$O$_8$ with $H \parallel c$ and compare the experimental
results to theoretical ones of a quasi-1D XXZ model studied by the stochastic series expansion (SSE)
quantum Monte Carlo simulations~\cite{Fan_PRR_2020}. The main results are summarized
in the phase diagram of Fig.~\ref{pd}. With the applied field, the Ising-antiferromagnetic (AFM) order is quickly suppressed.
But we do not resolve an LSDW order when further increasing the field. Surprisingly, over a
broad field range we identify a short-range transverse antiferromagnetic (SRTAF)
phase at high temperatures, whereas the magnetic moments
are found to be ordered and orientated along the
crystalline [110]/[1$\bar1$0] directions at low temperatures.
This SRTAF phase is likely to be stabilized by magnetic
impurities.
Fortunately, the TLL behavior survives in the
SRTAF phase, as evidenced by the power-law temperature
dependence of the spin-lattice relaxation rate $1/T_1$. The Luttinger
exponent $\eta$ is found to decrease from above 1 to below 1
when the field increases across about 7~T. This $\eta$ inversion suggests the dominant spin fluctuations change from longitudinal to transverse with increasing the field, which supports the
LSDW to TAF phase transition observed by neutron scattering~\cite{Shen_NJP_2019}.

The paper is organized as the following. Experimental methods
and numerical techniques
are presented in Sec.~\ref{stech}.
In Sec.~\ref{ssus}, we show the bulk susceptibility
data that exhibit the Ising AFM order and the TAF order
and compare the experimental data with the calculated Monte Carlo results.
In Sec.~\ref{sspec}, we report the NMR spectra
at typical fields and temperatures that evidence the high-temperature SRTAF phase. We then resolve
the magnetic patterns
by analyzing the NMR lineshape.
In Sec.~\ref{sslr}, we explore the low-energy excitations of the system from the $1/T_1$ data, and show the existence of
an $\eta$ inversion
at finite field. Based on these results, we propose
a phase diagram of the system and discuss the origin of the SRTAF phase
in Sec.~\ref{spd}.

\section{Techniques}
\label{stech}

The bulk susceptibility of SrCo$_2$V$_2$O$_8$ was measured in a Quantum-Design physical
property measurement system.
We performed NMR on $^{51}$V nuclei, which has the nuclear spin $I=7/2$ and the Zeeman factor $\gamma$=11.198~MHz/T.
Samples are cooled in a standard variable temperature insert for temperature above 1.5~K, and
in a dilution refrigerator for temperature at 0.5~K.
The spectra were obtained by the spin-echo technique with typical $\pi$/2 and $\pi$
pulses of 1$\mu$s and 2$\mu$s, respectively. For the broad spectra at low temperatures,
the entire spectra were obtained by
summing up the spectra with frequency sweeping.
The NMR Knight shift was calculated by $K_n=(f-{\gamma}H)/{\gamma}H$, where $f$
is the average frequency of the whole spectrum.

The hyperfine coupling among $^{51}$V nuclei and Co$^{2+}$ ions is primarily
of the pseudodipolar type~\cite{Ideta_PRB_2012,Kawasaki_JPS_2014,Cui_PRL_2019}.
Therefore, the effective dipolar field on a $^{51}$V nuclear spin, produced by Co$^{2+}$ moments, can be written as
\begin{equation}
{\vec H}_{dip}= \Sigma{_i}{{\mu_0}{[3(\vec\mu_i\cdot\vec r_i)\cdot \vec r_i-\vec\mu r_i^2}]/4{\pi}r_i^5},
\label{dipolar}
\end{equation}
where $\vec\mu_i$ is the moment of Co$^{2+}$ on site $i$, and $\vec r_i$ is the relative position
vector from the $i$-th Co$^{2+}$ to the $^{51}$V site. In this paper, in order to compare with
the NMR spectra in the ordered phase, we calculated the total hyperfine field
at each $^{51}$V site numerically by
summing over the dipolar field of Co$^{2+}$ moments
up to the fourth-nearest neighbors for accuracy.

The spin-lattice relaxation rate $1/T_1$ was measured
by the spin inversion-recovery method. $T_1$ was obtained by fitting
the nuclear magnetization $m(t)$ to the recovery function for
$I=7/2$ nuclei ~\cite{MacLaughlin_PRB__1971_4,Fujita1984},
$m(t)=m(\infty)-a[0.01191e^{-(t/T_1)^{\beta}} +0.06818e^{-(6t/T_1)^{\beta}}+0.20604e^{-(15t/T_1)^{\beta}}+0.71387e^{-(28t/T_1)^{\beta}}]$,
on the center (spin $-1/2{\rightarrow}1/2$) NMR line, summed over a frequency window of 50~kHz to reduce
the overlap with NMR satellites.
Here $\beta$ is a stretching factor, which was
found to be about 1 for temperatures above $T_N$. Note
that the above formula is accurate at high temperatures when the center line
is clearly distinguishable, but less accurate when the center line is partly mixed with
the neighboring NMR satellites (see Fig.~\ref{spec}).

To understand the experimental results, we consider a
model
defined on a 3D cubic lattice with the following
Hamiltonian
\begin{equation}\label{hamiltonian}
\begin{array}{l}
{\mathcal H} = {J_c}\sum\limits_i {{{\left( {{{\vec S}_i} \cdot {{\vec S}_{i + c}}} \right)}_\varepsilon }}  + {J_{ab}}\sum\limits_{i,\delta  = a,b} {{{\left( {{{\vec S}_i} \cdot {{\vec S}_{i + \delta }}} \right)}_{\varepsilon}}} \\
 - g{\mu _B}H\sum\limits_i {S_i^z},
\end{array}
\end{equation}
where
$\vec{S}_i$ is an $S{\rm{ = }}\frac{1}{2}$ spin operator at site $i$,
${J_c}$ is the nearest-neighboring intrachain exchange coupling,
${J_{ab}}$ is nearest-neighboring interchain exchange coupling, $g$ is the gyromagnetic ratio,
${\mu _B}$ is the Bohr magneton, and $H$ is the longitudinal magnetic field.
${\left( {{{\vec S}_i} \cdot {{\vec S}_j}} \right)_\varepsilon } \equiv \varepsilon \left( {S_i^xS_j^x + S_i^yS_j^y} \right) + S_i^zS_j^z$
is the XXZ type for nearest spin coupling, in which $\varepsilon$ is the anisotropy parameter.
The model is studied by using the SSE
quantum Monte Carlo simulations~\cite{Sandvik_PRE_2002}.
In our calculation, we
take $\varepsilon  = 0.5$, in consistency with the experimental data
~\cite{Kimura_JPSJ__2013,Bera_PRB__2014,Wang_PRB__2015,Wang_PRB__2016}, and ${J_{ab}} = 0.1{J_c}$
with the same anisotropic factor $\varepsilon$
for simplicity.
We set ${J_c}$ to be the energy unit and define the reduced
temperature $t = {T \mathord{\left/{\vphantom {T {{J_c}}}} \right.\kern-\nulldelimiterspace} {{J_c}}}$
and reduced field $h = g{\mu _B}{H \mathord{\left/{\vphantom {H {{J_c}}}} \right.\kern-\nulldelimiterspace} {{J_c}}}$.
The system size is up to $32 \times 32 \times 256$ and the lowest temperature calculated is $0.01J_c$.

\section{Bulk susceptibility and numerical simulations}
\label{ssus}

\begin{figure}[t]
\includegraphics[width=8.5cm]{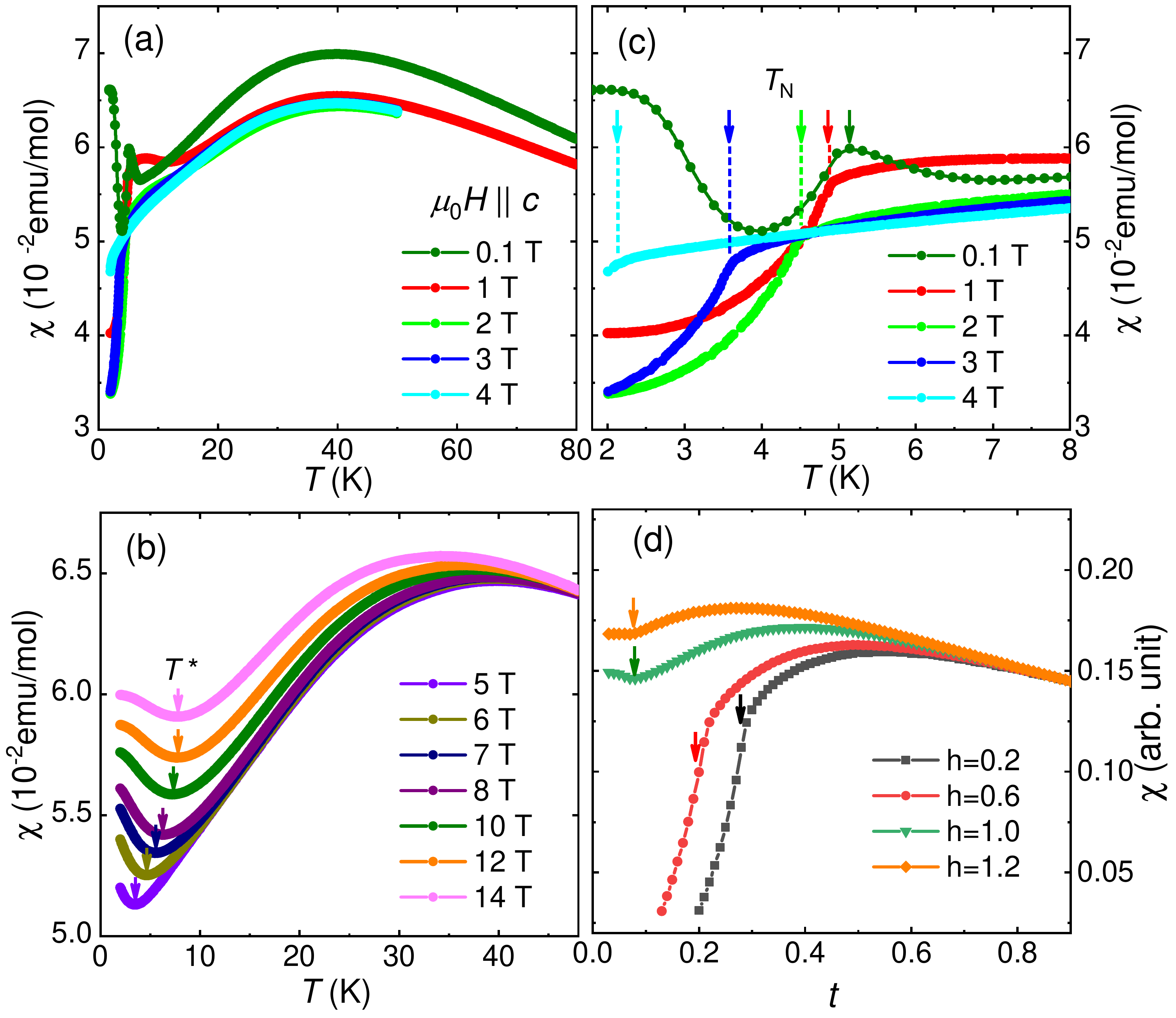}
\caption{\label{sus} (a)-(b) The magnetic susceptibility
versus temperature at fields from 0.1~T to 4~T and from 5~T to 14~T, respectively.
The arrows in (b) mark the onset temperature $T^*$ of the low-temperature
upturn in $\chi$($T$).
(c) An enlarged view of the low-field $\chi$($T$) near the
Ising-AFM transition.
Arrows mark the
transition temperature $T_N$.
(d) The magnetic susceptibility calculated by Monte Carlo simulations
on the quasi-1D model of Eq.~\eqref{hamiltonian}, where $t$ and $h$ are
reduced temperature and field in unit of the intrachain exchange coupling $J_c$, respectively.
The arrows mark the onset temperature of the downturn and the upturn in $\chi$($T$),
which characterize the transitions to the Ising-AFM and TAF phases, respectively.
}
\end{figure}

The dc susceptibility $\chi$($T$) measured under various fields
is shown in Fig.~\ref{sus}(a)-(c).
For all fields, $\chi$($T$) exhibits a hump feature at $T\approx$~40~K as
shown in Fig.~\ref{sus}(a)-(b).
This is a general feature of
the 1D magnetic systems and suggests that the exchange coupling is about 40~K~\cite{Bonner-Fisher}.
Now we focus on the evolution of $\chi$($T$) for temperatures below the hump.
For fields from 0.1~T to 4~T, $\chi$($T$) first decreases upon further cooling,
followed by a sharp drop below 5~K. This sharp drop is
caused by the Ising-AFM order, as the moments become ordered in parallel (or antiparallel) with the external field.
An enlarged view of the low-temperature $\chi$($T$) is further shown in Fig.~\ref{sus}(c),
where N\'{e}el temperature $T_N$ is determined by the sharp downturn temperature
in $\chi$ upon cooling, which gives from 5.1~K to 2~K, respectively,
with fields from 0.1~T to 4~T. These $T_N$s are further drawn in the phase diagram of
Fig.~\ref{pd}, which are consistent with those determined from elastic neutron scattering~\cite{Shen_NJP_2019}.
An upturn is seen at 0.1~T at temperatures below $T_N$ , which should be caused by
magnetic impurities. Similar behavior is also seen in the low-field
susceptibility with field applied along the $a$-axis~\cite{Cui_PRL_2019}.
With field from 1 to 4~T, the upturn behavior is no longer observed, indicating a
full polarization of impurity spins.

For fields from 5~T to 14~T, the sharp drop of $\chi$($T$) seen at low fields is absent, indicating
the absence of the Ising-AFM order at these fields, which is
consistent with the neutron scattering results~\cite{Shen_NJP_2019}.
Rather, an upturn in $\chi$($T$) at temperature $T^*$ below the hump is observed, as marked by the arrows in Fig.~\ref{sus}(b).
Since the upturn is not seen at low field, a Schottky contribution is ruled out.
With external field applied along the $c$-axis, this should attribute to a spin-flop-like
behavior seen by the bulk magnetization, that is, the TAF phase with short-range or
long-range ordered local moments on the $ab$ plane.

These downturn and upturn behaviors of susceptibility
associated with the respective Ising-AFM
and
TAF
orders
are verified by our
quantum Monte Carlo
calculations.
The calculated susceptibility ${\chi}=m/h$
versus the reduced temperature
at several different reduced fields is shown in Fig.~\ref{sus}(d).
For all the fields, there is clearly a hump feature at intermediate temperature.
At low fields
$h=$~0.2 and 0.6, the system is Ising AFM ordered at low temperatures, and
the sharp downturn in $\chi$ at low temperature
resolves the Ising-AFM ordering temperature
$T_N$,
consistent with the experiments. For $h=1.0$ and above, the system is in a TAF phase at low temperatures. The onset of the TAF order is identified by the upturn of $\chi$($T$)
as marked in Fig.~\ref{sus}(d).
This upturn feature is observed in the experimental susceptibility as shown above and
Knight shift data (shown later), and this confirms that the magnetic ordering for
fields at 5~T and above observed in experiments is of TAF type.
But different from the theoretical results, the observed $T^*$, the upturn temperature
in the high-field magnetic susceptibility in experiments,
is much higher than the ordering temperature of the low-field Ising antiferromagnetism. This comes from a disorder effect
which will be discussed in Sec.~\ref{sspec} with the NMR spectra.

\section{NMR spectra}
\label{sspec}

\begin{figure}
\includegraphics[width=8.5cm]{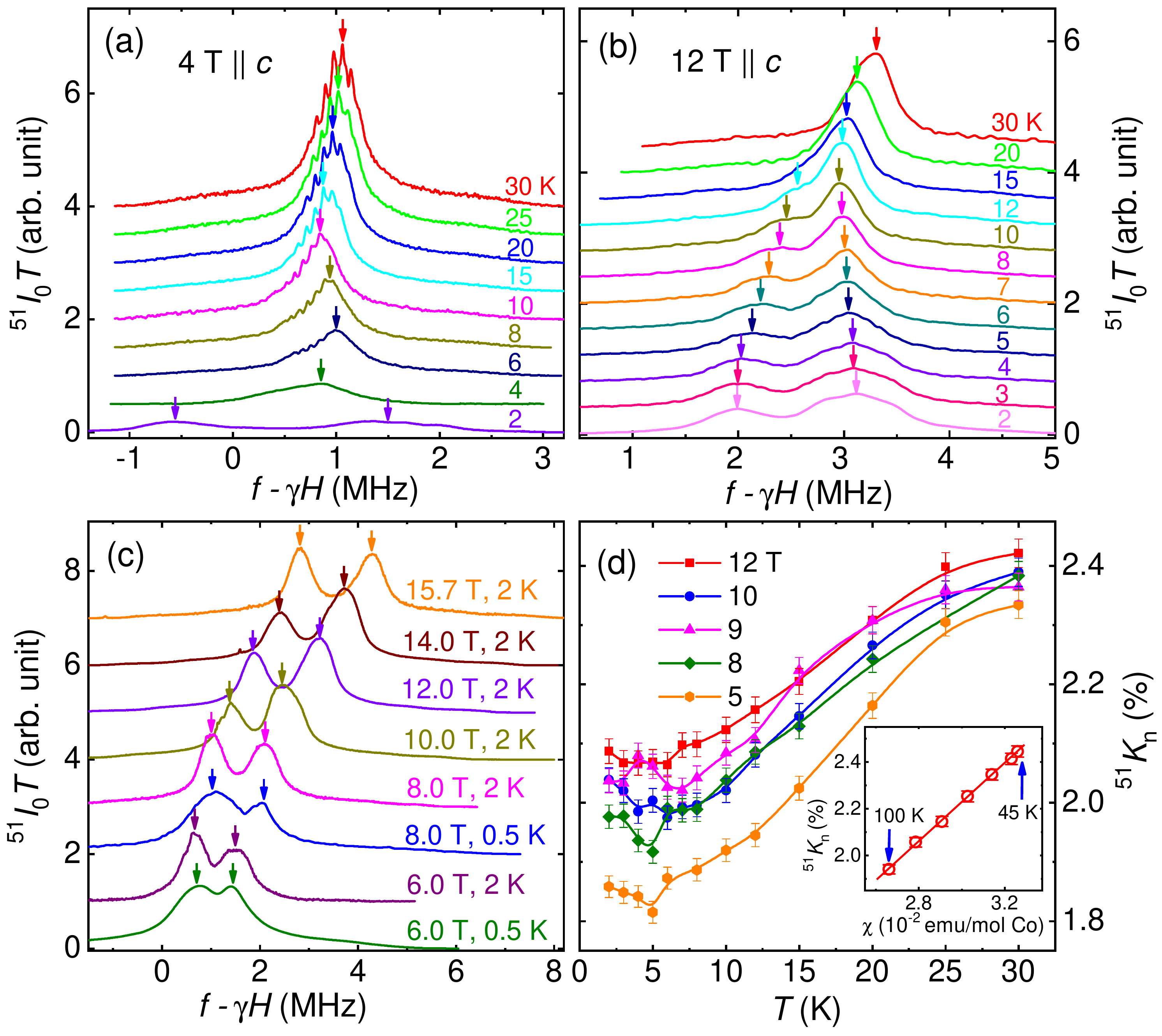}
\caption{\label{spec} (a)-(b) The $^{51}$V NMR spectra at typical temperatures, measured at 4~T and 12~T, respectively.
Vertical offsets are applied for clarity. Arrows mark the peak position of the center transition of the NMR lines.
(c) The NMR spectra at typical magnetic fields, measured at 2~K and 0.5~K.
(d) The Knight shift $^{51}K_n$ as functions of temperatures at typical fields. The inset: $^{51}K_n$
versus the magnetic susceptibility, $\chi$, measured at temperatures from 45~K to 100~K, under the same field of 5~T.
}
\end{figure}

\begin{figure*}
\includegraphics[width=13cm]{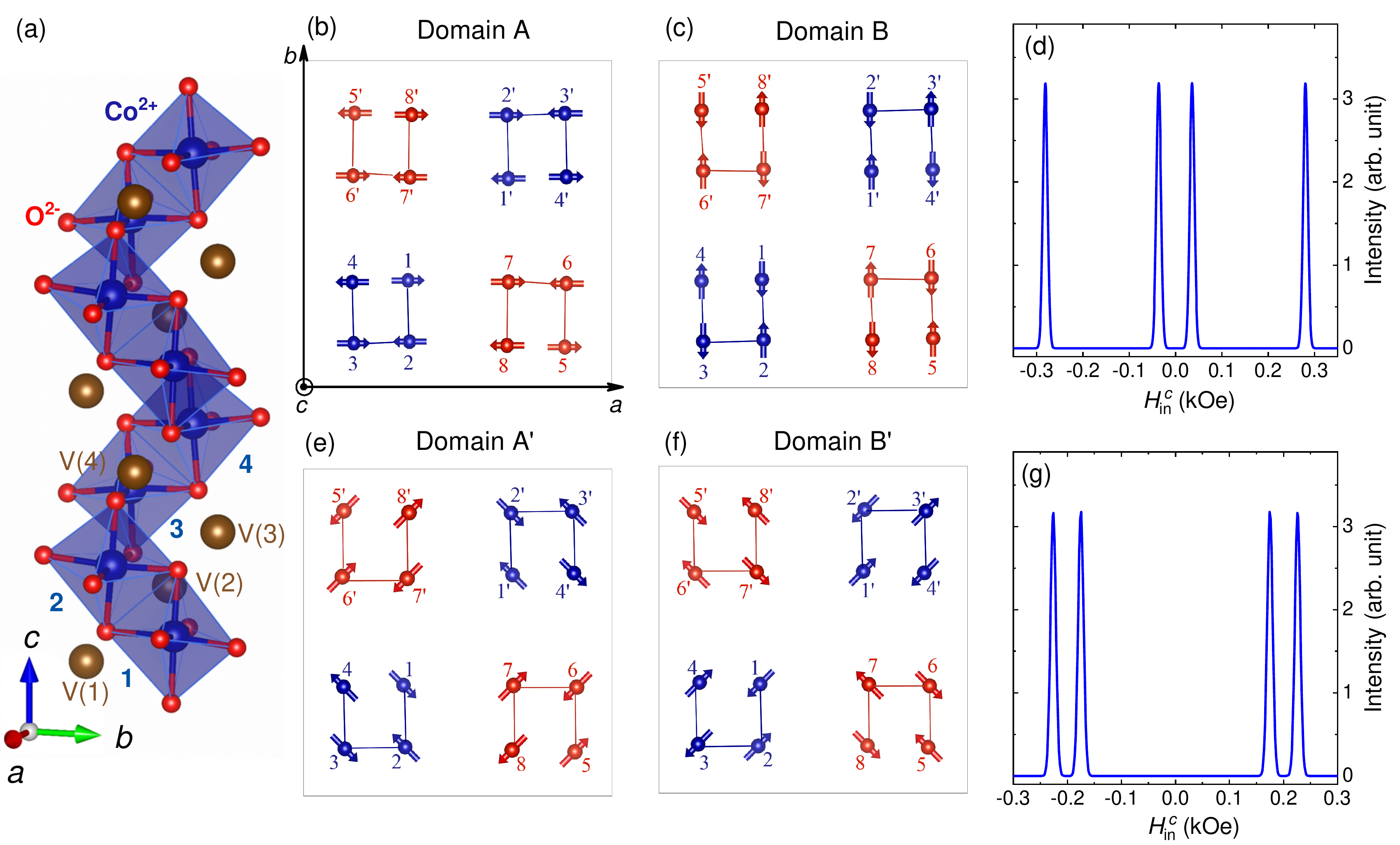}
\caption{\label{struc} (a) A sketch of a screw chain of SrCo$_2$V$_2$O$_8$.
V(1) to V(4) label four V sites on one chain in a unit cell.
(b)-(c) Illustration of the spin patterns in the TAF phase with moments orientate along the [100] or [010] direction, respectively, where domains A and B are twinned magnetic structures.
The numbers $1-8$ and $1'-8'$ label the sixteen Co$^{2+}$ ions within one unit cell, where red and blue colors
represent the two screwing directions along the chain. Arrows display the orientations of the local moments.
(d) Sketch of the ideal $^{51}$V NMR line calculated by
summing over contributions from domain A and domain B.
$H_{in}$ represents the total dipolar fields created by the Co$^{2+}$ moments on the $^{51}$V sites.
(e)-(f) Illustration of the spin patterns in the TAF phase with moments orientate along the [110] or [1$\bar{1}$0] directions, respectively with
twinned domains A$'$ and B$'$.
(g) Sketch of the ideal NMR line calculated by summing over the contributions from domains A$'$ and B$'$.
}
\end{figure*}

The $^{51}$V NMR spectra at several field values
with temperatures from 30~K down to 2~K are shown in Fig.~\ref{spec}(a)-(b).
At 4~T, seven transition lines with nearly equal frequency spacing $\nu_q{\approx}$~80~kHz are discernible at high temperatures. 
They correspond to one center line and six satellite lines caused by the coupling between the nuclear $^{51}$V quadrupole
moments and the local electric-field gradient.
Since $\nu_q$ is very small compared to the Zeeman frequency ${\gamma}H$,
it impacts the frequency of the center NMR line and
the Knight shift by a second-order correction of ${\nu_q}^2/{\gamma}H$, which
is almost negligible.

At 4~K, a peaked NMR spectrum, where the center and satellite transitions overlap due to line broadening,
is seen, showing the system is in the paramagnetic phase~\cite{Kuo_SSC_2009,Kawasaki_JPS_2014}.
At and below 2~K, the NMR line splits into two peaks with a frequency separation of about 2.2~MHz.
The two peaks are at negative and positive frequencies to $\gamma H$, respectively, with nearly equal spectral weight.
This is a clear evidence for the Ising-antiferromagnetic order, which generates
negative and positive hyperfine fields, respectively.
The transition temperature to the Ising ordered phase is determined to be about 2.3~K
from the $1/T_1$ measurement at this field (see Fig.~\ref{slrr}),
which is also consistent with neutron scattering studies~\cite{Shen_NJP_2019}.

At high fields, surprisingly, a two-peak feature also develops in the NMR line even at high temperatures.
As shown at 12~T, the two-peak feature is already seen when cooled below 12~K [Fig.~\ref{spec}(b)],
which is well above $T_N$ as revealed by neutron scattering measurements
at the same field~\cite{Shen_NJP_2019}.
Indeed, the frequency separation of two peaks increases
upon cooling, with the left peak shifting toward lower frequency,
and the right peak shifting toward higher frequency [Fig.~\ref{spec}(b)].
Similar line splits are observed in a range of magnetic fields,
as shown in Fig.~\ref{spec}(c). The $^{51}$V spectra at 0.5~K and 2~K,
with fields from 6~T to 15.7~T, all exhibit the two-peak feature.
This, as will be discussed later, is
associated with an emergent TAF order, either short ranged or long ranged.

Note that in the paramagnetic states, with external field applied along the $c$ axis, all the $^{51}$V
in the unit cell should be equivalent, which leads to a single NMR center line as
observed at 30~K. The low-temperature line splitting
then indicates the onset of a static magnetism. For the Ising-AFM or the LSDW states,
the spectrum should contain both negative and positive frequencies relative to ${\gamma}H$, as shown
with field at 4~T in Fig.~\ref{spec}(a). However, this is not what was observed at 6~T and higher fields.
Rather, both NMR peaks locate at positive frequencies.
This is instead consistent with
a local TAF order where the ordered
local moments orientate on the $ab$ plane as discussed below.

At high fields, the magnetic structure should be characterized
by the TAF type with ordered moments lying on the $ab$ plane.
Based on the neutron scattering data,
two magnetic structures for the TAF phase
have been proposed but not yet resolved:
The moments orientate either along the [100]/[010] directions as a collinear type,
or along the [110]/[1$\bar{1}$0] directions as a noncollinear type~\cite{Grenier_PRB_2015}.
For each proposed structure, doubly degenerate magnetic domains
can coexist, which are labeled as domain A and B
for the collinear type as shown in Fig.~\ref{struc} (b)-(c),
or domain A$'$ and B$'$ for the noncollinear type as shown in Fig.~\ref{struc}(e)-(f).


We then try to determine the local magnetic structure by simulating the NMR spectra
from the two proposed magnetic configurations.
We calculated the dipolar field on 16 $^{51}$V sites with their coordinates,
contributed from Co$^{2+}$ sites up to the fourth-nearest neighbors
with the formula shown in Eq.~(1).
We assume temporarily that staggered, planar antiferromagnetic moment is 1~$\mu_B$/Co$^{2+}$
for the AFM phases shown in Fig.~\ref{struc}.
The calculated dipolar fields along the $c$ direction, $H_{\rm in}^c$, for the
collinear-type are listed in Table~I(a), and for the noncollinear-type
are listed in Table~I(b). Four types of $H_{\rm in}^c$ are obtained for each type,
which indicates that four inequivalent V sites. We labeled the inequivalent
V sites as V(1), V(2), V(3) and V(4), also shown in Fig.~\ref{struc}(a).

The $^{51}$V spectra are then simulated for the collinear-type structure as sketched in Fig.~\ref{struc}(d),
and for the noncollinear-type structure as sketched in Fig.~\ref{struc}(g).
Note the uniform magnetization of Co$^{2+}$ along the $c$ axis will only cause a bulk
shift of the whole spectra, which is not included in the calculation.
Domains A and B produce identical NMR line shapes, also for domains A$'$ and B$'$.
The simulated spectra of both structures contain four NMR lines but
with very different frequency separations. For the collinear
structure, the NMR spectrum contains
one line at a negative frequency, two close lines at
intermediate frequencies, and one line at a positive frequency.
By contrast, for the noncollinear structure, one pair of lines
is located at negative frequencies, and the other pair of lines
is at positive frequencies.

The measured spectra shown in Fig.~\ref{spec}(b)-(c) are more consistent with
the spectral profile of the noncollinear
magnetic structure when we consider the broadening of the two lines within each pair which
could not be resolved experimentally, if disorder
broadens each NMR line significantly.
The difference of the hyperfine fields
of two NMR peaks is about 0.4~kOe. By contrast, the experimental line split at 2~K and 12~T
[Fig.~\ref{spec} (b)], for example, is as large as 1.1~kOe.
In fact, two factors should be considered on these quantitative difference.
First, the actual dipolar field is proportional to the ordered moment of Co$^{2+}$,
relative to the current assumption of 1~$\mu_B$/Co$^{2+}$.
Second, the off-diagonal hyperfine field should be much larger with the pseudodipolar type than that with the
pure dipolar type as suggested by previous works
in BaCo$_2$V$_2$O$_8$~\cite{Kawasaki_JPS_2014} and SrCo$_2$V$_2$O$_8$~\cite{Cui_PRL_2019}.

\begin{table}
\centering
\begin{tabular}{c|m{4.5cm}<{\centering}|c|c}
\multicolumn{2}{l}{(a) Collinear type}\\
\hline
\multirow{2}*{~Site~}&\multirow{2}{3cm}{Fractional coordinates (x, y, z) (n, m = 0 or 1)}&\multicolumn{2}{c}{$H_{\rm in}^c$ (kOe)}\\
\cline{3-4}
~& ~ & Domain A & Domain B \\
\hline
  V(1) &  (0.2608-0.0216n+0.5m, 0.08+0.5n+0.34m, 0.0934) & -0.0361  & -0.2809 \\
\hline
  V(2) &  (0.08+0.34n+0.5m, 0.2392+0.5n+0.0216m, 0.3434) & -0.2809 & 0.0361   \\
\hline
  V(3) & (0.7392+0.0216n-0.5m, 0.08+0.5n+0.34m, 0.5934) & 0.0361 & 0.2809 \\
\hline
  V(4) &  (0.92-0.34n-0.5m, 0.2392+0.5n+0.0216m, 0.8434) & 0.2809 &  -0.0361  \\
\hline
\end{tabular}
\vskip3mm
\begin{tabular}{c|m{4.5cm}<{\centering}|c|c}
\multicolumn{2}{l}{(b) Noncollinear type}\\
\hline
\multirow{2}*{~Site~}&\multirow{2}{3cm}{Fractional coordinates (x, y, z) (n, m = 0 or 1)}&\multicolumn{2}{c}{$H_{\rm in}^c$ (kOe)}\\
\cline{3-4}
~& ~ & Domain A$'$ & Domain B$'$ \\
\hline
  V(1) &  (0.2608-0.0216n+0.5m, 0.08+0.5n+0.34m, 0.0934) & -0.2261  & -0.1749 \\
\hline
  V(2) &  (0.08+0.34n+0.5m, 0.2392+0.5n+0.0216m, 0.3434) & -0.1749 &  0.2261    \\
\hline
  V(3) & (0.7392+0.0216n-0.5m, 0.08+0.5n+0.34m, 0.5934) & 0.2261 & 0.1749  \\
\hline
  V(4) &  (0.92-0.34n-0.5m, 0.2392+0.5n+0.0216m, 0.8434) & 0.1749 & -0.2261 \\
\hline
\end{tabular}
\caption{(a)-(b) The calculated $c$-axis component of the dipolar field,
$H_{\rm in}^c$, on 16 $^{51}$V nuclei in a unit cell, each summed over contributions from four
neighboring Co$^{2+}$ spins (see the text), for the collinear type and noncollinear type respectively.
}
\label{tab}
\end{table}

We further examine the Knight shift $^{51}K_n$ data.
$^{51}K_n$ is calculated from the average frequency of the
spectrum, and shown as a function of temperature
in Fig.~\ref{spec}(d) for fields from 5~T to 12~T.
Note that the Knight shift is related to the bulk susceptibility by
$K_n = A_{hf}\chi/N_A\mu_B$, where $N_A$ is the Avogadro constant, $\mu_B$ is the Bohr magneton,
and $A_{hf}$ is the hyperfine coupling constant.  In the inset of Fig.~\ref{spec}(d),
$^{51}K_n(T)$ at 5~T is then plotted against $\chi$($T$), which 
is measured with temperature from
45~K to 100~K (data not shown). A linear scaling is clearly seen, and
the slope of the line gives $A_{hf}{\approx}$~0.46~T/$\mu_B$.

For fields above 5~T, $^{51}K_n$ first decreases with
temperature when cooled below 30~K, and then shows an upturn
below about 5~K. This upturn behavior is also
observed by the bulk susceptibility  $\chi$($T$), as shown earlier.
The higher onset temperature of the NMR line split than that of the upturn in the bulk susceptibility can be understood by
the development of short-range order at temperatures above $T^*$,
because the uniform magnetization, revealed by the bulk susceptibility,
still decreases upon cooling. Whereas below $T^*$, a LRTAF
tends to be developed with the onset of susceptibility upturn and
the constant NMR line splits with temperatures down to 0.5~K~[Fig.~\ref{spec}(c)].

We therefore conclude that
a short-range order with the noncollinear
antiferromagnetism, as shown in Fig.~\ref{struc}(e) and (f),
is established in SrCo$_2$V$_2$O$_8$ far above $T_N$. Note that the onset of the
short-range order was not reported in BaCo$_2$V$_2$O$_8$~\cite{Klanjsek_PRB__2015}, which may suggest that
the phase is sensitive to disorder, a detailed discussion of which will be given in Sec.~\ref{spd}.

\section{the spin-lattice relaxation rate}
\label{sslr}

In general, the spin-lattice relaxation rate $1/T_1$ probes low-energy fluctuations,
with $1/T_1$ $\sim$ $T\sum_{q}A_{hf}(q) Im\chi^{+-}(q,\omega)/\omega$ for
magnetic systems, where
$A_{hf}$ is hyperfine coupling constant, $\chi^{+-}$ is the dynamic susceptibility of the
system, and $\omega$ is the Larmor frequency of the nuclei.

\begin{figure}
\includegraphics[width=8.5cm]{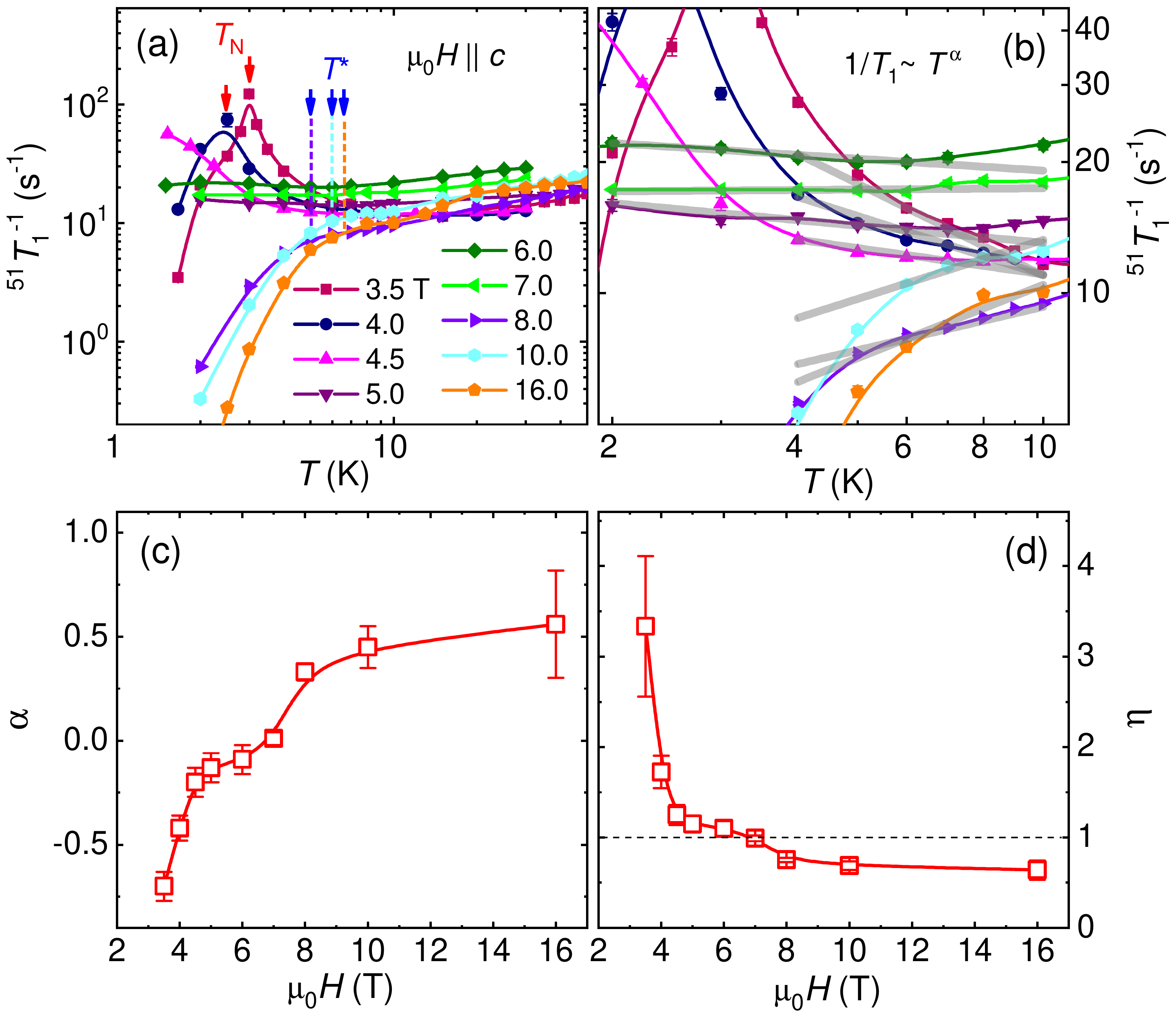}
\caption{\label{slrr} (a) The spin-lattice relaxation rate $1/^{51}T_1$ as functions of
temperatures at various fields. The red arrows mark the peak position, which determines $T_N$ at low fields.
The blue arrows mark the downturn temperature of $1/^{51}T_1$, labeled as $T^*$.
(b) An enlarge view of the low-temperature $1/^{51}T_1$ data. Gray lines are the fits of data
to the power-law function $1/T_1\sim T^\alpha$ for temperatures above $T_N$, under different fields.
(c) The power-law exponent $\alpha$ as a function of the field obtained from the
fitting.
(d) The Luttinger exponent $\eta$ as a function of the field. The dotted line
corresponds to $\eta=1$. It intersects with
the $\eta$ curve at about 7~T, indicating an $\eta$ inversion.
}
\end{figure}

For this compound, we report the spin-lattice relaxation rate $1/^{51}T_1$
measured on the only NMR peak at high temperatures,
and the right peak in the SRTAF phase (see Fig.~\ref{spec}).
$1/^{51}T_1$ as functions of temperature is shown in Fig.~\ref{slrr}(a),
for fields from 3.5~T to 16~T.
Note that the $1/^{51}T_1$ on the left peak (data not shown)
is about 10$\%$ lower and follows the same temperature dependence as the right one.

\begin{figure}
\includegraphics[width=8.5cm]{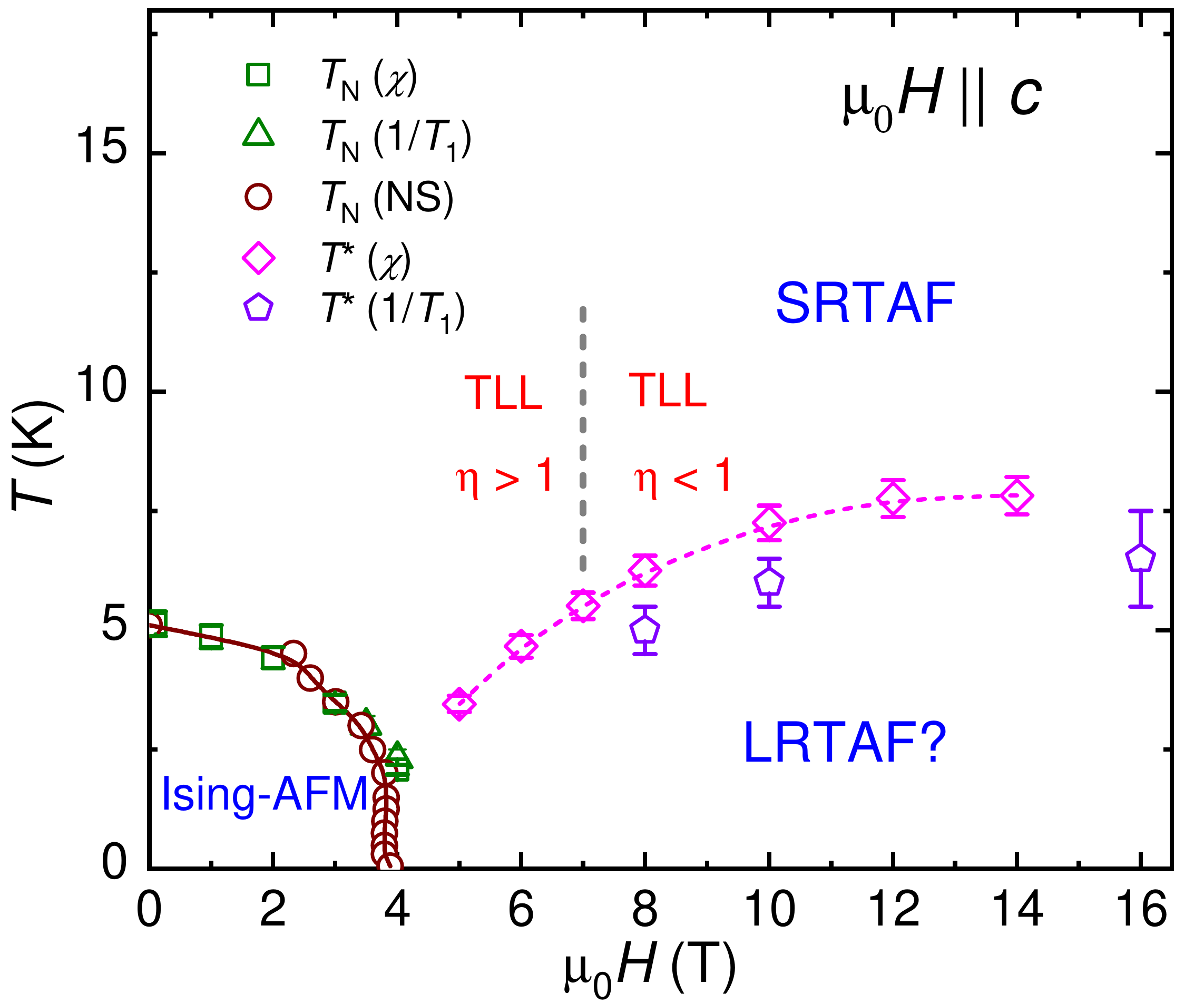}
\caption{\label{pd} The phase diagram of SrCo$_2$V$_2$O$_8$ 
under a longitudinal magnetic field.
Squares and triangles represent the $T_N$
determined by the susceptibility and the spin-lattice relaxation rate measurements, respectively.
The brown circles are the $T_N$ determined
from neutron scattering,
adapted from Ref.~\citenum{Shen_NJP_2019}.
Pink diamonds represent $T^*$, the upturn temperature of magnetic susceptibility,
and violet diamonds represent the downturn temperature of $1/^{51}T_1$, also labeled as $T^*$.
The dotted gray line shows the $\eta$ inversion in the TLL.
The position of the label SRTAF denotes the region below which the NMR line split, whereas
a LRTAF may appear below $T^*$.
}
\end{figure}

At 3.5~T, $1/^{51}T_1$ first decreases when cooled from 30~K down to 10~K,
followed by an upturn and a sharp peak at 3~K, which evidences the
Ising-AFM transition at $T_N$
with
enhanced low-energy spin fluctuations.
At 4~T, a transition temperature $T_N$=2.3~K is also resolved clearly.
The
determined $T_N$ values at different fields are then plotted in Fig.~\ref{pd},
which
show a monotonic decrease with the field,
consistent with the $\chi$($T$) and the neutron scattering data~\cite{Shen_NJP_2019}.
For fields from 5~T to 7~T, $1/^{51}T_1$ barely change when
cooled from 10~K to 1.5~K, which suggests that the system is close
to a magnetic quantum critical point.
For fields at 8~T and above, $1/^{51}T_1$ drops when cooled below 20~K; below about 7~K, a
very rapid drop of $1/^{51}T_1$ is seen, which probably indicates a crossover to LRTAF phase.
The downturn temperature of $1/^{51}T_1$ for different fields is
labeled as $T^*$ and
the $T^*$ line is plotted in the phase diagram in Fig.~\ref{pd}.
This $T^*$ is slightly lower by about 1.3~K than
that determined by the susceptibility upturn.

An enlarged view of the low-temperature data is shown in
Fig.~\ref{slrr}(b) from 2~K to 10~K.
In the temperature regime just below 10~K,
the $1/^{51}T_1$ data can fit to a power-law function
$1/^{51}T_1 \sim T^\alpha$, with varying exponent $\alpha$, as shown in Fig.~\ref{slrr}(b).
Note that this power-law behavior is different from the critical enhancement
when the system is approaching the Ising-AFM transition,
but is a signature of the
1D spin fluctuations
existing well above the magnetic ordering temperature~\cite{Jeong_PRL_2016}.
$\alpha$, obtained from the power-law fitting, is then
shown in Fig.~\ref{slrr}(c) as a function of the field.
It increases monotonically with the field from $-0.7$ at 3.5~T to 0.56 at 16~T.

When the Ising-AFM order is suppressed by the field, the system becomes gapless. At low temperatures,
an LSDW or a TAF order can be stabilized due to the interchain couplings.
In the paramagnetic phase where these orders are destroyed by increasing
the temperature, the system is described by a TLL~\cite{Haldane_PRL_1980, Kimura_PRL_2008}.
In a TLL, the spin-spin correlation function $C(r)$
exhibits a power-law decay with the distance $r$ between the spins,
$C(r){\sim}r^{-\eta}$, where $\eta$ is the Luttinger exponent.
For an Ising anisotropic spin chain system under a longitudinal field, when the Ising order
is suppressed by the field, the system is first driven to an LSDW ground state. At finite temperature where the LSDW order is melted,
longitudinal spin fluctuations are dominant with $\eta>1$
~\cite{Klanjsek_PRL_2008,Bouillot_PRB_2011}. With further increasing the field, the ground state changes to a TAF phase. As a consequence, transverse spin fluctuations become dominant
with $\eta<1$ in the finite-temperature paramagnetic phase~\cite{Klanjsek_PRL_2008,Bouillot_PRB_2011}.
The dominant spin fluctuations in the TLL can be detected from the temperature dependence of the NMR $1/T_1$:
$1/{T_1}\approx c{T^{1/\eta- 1}}$ if the main contribution comes from longitudinal correlations
and
$1/{T_1}\approx c{T^{\eta- 1}}$ if the contribution is mainly from the transverse correlations~\cite{Klanjsek_PRB__2015}.

We can extract the Luttinger exponent $\eta$ from the power-law temperature dependence of the $1/{T_1}$ data.
Our results suggest the $1/T_1$ for this compound mainly detects the longitudinal spin correlations, and this leads to $\alpha=1/\eta-1$.
The contribution from transverse fluctuations,
however, is minor. This could be either because of the particular hyperfine coupling structure of this compound, or
due to
the onset of SRTAF order, which suppresses the transverse fluctuations.
The values of $\eta$ evolved with the field, calculated from $\alpha$,
are shown in Fig.~\ref{slrr}(d). An $\eta$ inversion where $\eta$ decreases
monotonically from  $\eta > 1$ to $\eta < 1$ with increasing the field is observed
at about 7~T. This $\eta$ inversion suggests that
the dominant spin fluctuations change from longitudinal to
transverse when the field is increased across 7~T~\cite{Okunishi_PRB_2007},
and is therefore consistent with the change of the ground state from the
LSDW phase to the TAF phase at 7~T observed in a recent neutron scattering study~\cite{Shen_NJP_2019}.
Similar $\eta$ inversion behavior determined from the $1/T_1$ measurements has been reported
in BaCo$_2$V$_2$O$_8$~\cite{Klanjsek_PRB__2015,Horvatic_arxiv_2020}
and predicted theoretically in YbAlO$_3$~\cite{Fan_PRR_2020}.

\section{Phase diagram and discussions}
\label{spd}

Our results are summarized
as the phase diagram in the $H$-$T$ parameter space shown in Fig.~\ref{pd}.
The $T_N$ of the Ising-AFM order
below 4~T determined from several different probes are consistent.
For fields above 4~T, we do not observe
an LSDW order. However, the NMR line splits and the upturn in the magnetic susceptibility
suggest that a possible LRTAF develops at a temperature much higher than that reported by
the neutron scattering~\cite{Shen_NJP_2019}.
Furthermore, we find a SRTAF phase over a broad field and temperature regime above $T^*$,
which survives as high as
12~K at 12~T, by the clear evidence of NMR line splits as shown in Fig.~\ref{spec}(b).
Our findings of missing the long-range LSDW and
the high onset temperature of SRTAF and possible LRTAF order
are in direct contrast to the results of a recent neutron scattering
measurement~\cite{Shen_NJP_2019}, in which LSDW and TAF phases were
revealed to be stabilized below 1.5 K sequentially with field~\cite{Shen_NJP_2019}.

A most likely explanation is that the disorder effect in our sample
is stronger.
Given that the impurity spins are fully polarized with field
above 1~T as observed above, they orientate along the crystalline
$c$ direction with applied field. Such random moments should
suppress the LSDW order with the same orientation. However, for the TAF phase,
the magnetic ordering may not be suppressed,
as the impurity moments are perpendicular to the Co$^{2+}$ moments.
This explains why the LSDW order is not seen in our samples.
Furthermore, although the disorder usually disturbs the magnetic order
in unfrustrated systems, in frustrated systems, it can help stabilizing
the magnetic order via relieving the frustration as described below.

Note that frustrated interchain couplings are
important and necessary to stabilize the pattern of
the Ising antiferromagnetic order~\cite{Niesen_PRB_2013,Niesen_PRB_2014,Klanjsek_PRB__2015},
as well as the noncollinear
spin patterns~\cite{Grenier_PRB_2015} shown in Fig.~\ref{struc}(e) and (f).
The frustrated interchain couplings provide an important clue to
understand the discrepancy between our results and those of Ref.~\citenum{Shen_NJP_2019}.
In the quasi-1D spin systems, the magnetic impurities
may effectively increase the AFM interchain
coupling by blocking the frustrated exchange paths. Furthermore,
the anisotropy of interchain coupling may also be affected by disorder, given the very anisotropic crystal field environment
of the compound.
To understand this, we performed Monte Carlo simulations
by reducing the anisotropy of the interchain coupling, while the intrachain
coupling remains the same. We found that by reducing the Ising anisotropy,
the $T_N$ of the Ising antiferromagnetism
is only slightly
suppressed, whereas the  $T_N$ of the LRTAF is strongly enhanced.
This may stabilize the LRTAF at low temperature and pins the short-range fluctuations at high
temperatures. These features are
exactly what we observed in our experiments.

Interestingly, though the long-range LSDW is not stabilized at low temperatures,
the TLL behavior still emerges at low energies,
regardless of the existence of SRTAF phase.
The TLL behavior still appears at about 3.5~T in the paramagnetic phase, and
the dominant spin fluctuations switch
from the longitudinal type to the transverse type, as evidenced by the $\eta$ inversion effect
observed at about 7~T. Note that the $\eta$ inversion occurs at exactly the same field
where a transition from the LSDW ground state to the TAF
one was observed in the neutron scattering~\cite{Shen_NJP_2019},
and similar behavior is also observed in BaCo$_2$V$_2$O$_8$~\cite{Grenier_PRB_2015,Klanjsek_PRB__2015}.
Although the long-range LSDW is destructed and a pinned short-range
ordered TAF phase emerges in our sample by the disorder effect,
the fingerprints of dominant longitudinal spin fluctuations
at intermediate fields are still observed.
Coexistence of longitudinal and transverse spin fluctuations
was also suggested by the independent terahertz~\cite{Wang_nature_2018} and
neutron scattering~\cite{Bera_NP_2020} measurements.

\section{Summary}

To summarizy, we report the magnetic phase diagram of the AFM screw chain compound SrCo$_2$V$_2$O$_8$ under
a magnetic field applied along the $c$ axis by performing the NMR and magnetic
susceptibility measurements and supplemented by quantum Monte Carlo simulations.
At low fields, the system exhibits an Ising-AFM order below about 5~K.
This Ising-AFM order is quickly suppressed by the magnetic field.
For fields at 5~T and above, no long-range LSDW order is detected.
However, the upturn of $\chi$ and $^{51}K_n$, as well as the
NMR line splitting, indicate the onset of a short-range TAF phase  even at very high temperatures,
and a possible long-range TAF order with enhanced transition temperatures.
We further show that the spins in the TAF phase form a noncollinear-type pattern,
with moments aligned along the crystalline [110]/[1$\bar1$0] directions.
This SRTAF order is likely pinned by magnetic impurities in the compound,
which suggest an anomalous sensitivity of the phase to the interplay of magnetic impurities
and interchain couplings.
Nevertheless, the power-law temperature dependence of the $1/^{51}T_1$ data reveals that the low
energy spin dynamics in the paramagnetic phase follow the TLL behavior, survived from the SRTAF order.
We extract the Luttinger exponent $\eta$, and show that there is an $\eta$ inversion
at about 7~T, across which the dominant spin fluctuations of the system
turn from the longitudinal to the transverse type.

\section*{Acknowledgments}

We thank Prof. Bruce Normand for helpful discussions.
This work was supported by the National Natural Science Foundation of China
under Grants No.~12104503, No.~12134020, No.~12174441, and No.~51872328, the
Ministry of Science and Technology of China under Grant No.~2016YFA0300504,
the China Postdoctoral Science Foundation under Grant
No.~2020M680797, the Fundamental Research Funds for the Central
Universities and the Research Funds of Renmin University of China
under Grants No.~21XNLG18, No.~20XNLG19, and No.~18XNLG24.

\end{document}